# Focusing of light energy inside a scattering medium by controlling the time-gated multiple light scattering

Seungwon Jeong[1,2,+], Ye-Ryoung Lee[1,2,+], Sungsam Kang[1,2,+], Wonjun Choi[1,2], Jin Hee Hong[1,2], Jin-Sung Park[2], Yong-Sik Lim[3], Hong-Gyu Park[2], and Wonshik Choi[1,2,*]

[1]*Center for Molecular Spectroscopy and Dynamics, Institute for Basic Science, Seoul 02841, Korea*
[2]*Department of Physics, Korea University, Seoul 02841, Korea*
[3]*Department of Nano Science and Mechanical Engineering and Nanotechnology Research Center, Konkuk University, Chungbuk 380-701, Korea*

Abstract: The efficient delivery of light energy is a prerequisite for non-invasive imaging and stimulating of target objects embedded deep within a scattering medium. However, injected waves experience random diffusion by multiple light scattering, and only a small fraction reaches the target object. Here we present a method to counteract wave diffusion and to focus multiple-scattered waves to the deeply embedded target. To realize this, we experimentally inject light to the reflection eigenchannels of a specific flight time where most of the multiple-scattered waves have interacted with the target object and maximize the intensity of the returning multiple-scattered waves at the selected time. For targets that are too deep to be visible by optical imaging, we demonstrated a more than 10-fold enhancement in light energy delivery in comparison with ordinary wave diffusion cases. This work will lay a foundation for enhancing the working depth of imaging, sensing, and light stimulation.

**Introduction**

Objects of interest are often embedded within disordered environments in many important *in vivo* applications such as biomedical imaging[1], phototherapy[2], and optogenetics[3]. In these applications, it is necessary to deliver light waves to the deeply embedded target object for efficient optical imaging, sensing, and light stimulation. However, random wave diffusion induced by multiple light scattering on these disordered environments drastically limits the ability to reach the object of interest. When waves are propagating inside a scattering medium, they are spread in both space and time and only a small fraction of the injected energy reaches the target object. A simple solution would be to increase the injecting energy, but this will increase the background noise and induce unwanted damage to the sample. To extend the working depth of optical methodologies, it is necessary to develop methods that increase the efficiency of energy delivery to the embedded target.

In the past decades, numerous studies demonstrated the control of light waves traversing a scattering medium. The underlying concept is to control the wavefront of illumination light to tailor the interference of multiple-scattered waves. In a pioneering work by Vellekoop et al., a light wave transmitted through a scattering layer was focused by the wavefront shaping of an incident wave[4]. Afterwards, temporal as well as spatial focusing has been realized by using broadband light sources[5-7]. Studies have also been conducted to enhance the total energy transmitted through a scattering medium. In these studies, a scattering matrix was measured and its eigenchannels were identified. Light waves coupled to the eigenchannels with large eigenvalues were shown to have extraordinary transmittance[8-10]. While all these studies have shown great potential to exploit multiple light scattering, the target point for which light waves are controlled is located outside of the scattering medium and not inside. Therefore, their practical use is limited for *in vivo* applications since the methodologies cannot aim at the embedded targets.

Adaptive optics is one of the most representative approaches concerning the wave control within a scattering medium. It corrects the sample-induced phase retardations of the so-called ballistic photons or single-scattered waves by using a wavefront shaping device[11-15]. Coupling light to the reflection eigenchannels in the steady-state measurements acts similar to the adaptive optics[16]. However, the adaptive optics can control only a tiny fraction of the internal waves for a deeply embedded target because the single-scattered waves are orders of magnitude weaker than the multiple-scattered waves. Time-reversal of the acoustically modulated optical waves is another approach that enables focusing optical waves to the acoustic focus within the medium[17,18]. However, the concept was demonstrated using the transmitted wave through scattering medium, and further studies are needed to use phase-conjugation of the backscattered waves to warrant its utility for *in vivo* applications. Feedback control of the wavefront to increase the fluorescence signal from fluorophores embedded within a scattering layer is another approach to enhance light energy delivery to the target, but this requires labeling agents[19]. In addition, the operation is not target-specific since all the fluorophores within the sample are affected.

Here, we present the first experimental implementation of the time-gated reflection eigenchannels for increasing the intensity of backscattered waves arriving only at a selected flight time. By tuning the flight time to that determined by the depth of the target object, we could suppress random wave diffusion inside the scattering medium and focus light waves, especially multiple-scattered waves, to a target object that is embedded too deep to be resolved by optical imaging. In doing so, we send over 10 times more light energy to the target object in comparison with uncontrolled inputs. The method worked for target objects embedded at a depth of approximately 2 times the transport mean free path, at which target objects are completely invisible by optical imaging. We demonstrated the enhanced energy delivery through a rat's skull as well as through artificial scattering samples. The proposed method is not invasive at all as it controls back-scattered waves in the epi-detection geometry. Therefore, it can potentially be employed to many important applications in life science and biomedicine where conservation of the sample's original conditions is critical.

**Principle**

Let us consider a typical sample geometry in which a target object is located inside a scattering medium (Fig. 1a). For a pulsed wave incident to this sample, the intensity of backscattered waves can have a flight time-dependent profile as shown by the dashed blue curve in Fig. 1b. Depending on the flight time, we can categorize the detected backscattered waves into four representative trajectories, labeled as (1) - (4) in the order of increasing flight time: Trajectory (1) indicates back-reflection at the surface of the scattering medium. Trajectory (2) contains multiple-scattered waves whose flight times are shorter than the flight time $\tau_0 = 2z_s/c$ set by the target depth $z_S$ and the average speed of light $c$ in the medium. Trajectory (3) includes single-scattered waves or ballistic photons that are scattered only one time by the target object, but not at all by the scattering media. The flight time of these single-scattered waves is set by $\tau_0$. Due to the finite width of the time-gating window, the so-called 'snake-like' multiple-scattered waves[20] that have experienced highly forward scattering are detected at the same time gating window. Some of the snake-like waves interact with the target object and can thus be regarded as signal waves, while the others act as noise. In fact, the snake-like signal waves are far stronger than the single-scattered waves for deeply embedded targets (see Supplementary Section I for numerical analysis). Therefore, it is important to control the snake-like multiple-scattered waves for an efficient light energy delivery to the target. Trajectory (4) represents multiple-scattered waves that mostly interact with the scattering medium and travel deeper than $z_S$. In our approach, we apply the time-gated detection to selectively detect trajectory (3) that contains signal waves and to rule out the background scattering from the trajectories (1), (2) and (4). Because of this background noise rejection, the sensitivity of time-gated detection is much higher than that of the steady-state measurements where all the backscattered waves are detected at once.

An important next step that we propose is to control wave propagation in a way that allows to maximize the intensity of light waves traveling via trajectory (3) (red curve in Fig. 1b). This is realized by injecting light to the unique eigenchannels of the given scattering medium that have an extraordinarily large reflectance at the flight time $\tau_0$. These eigenchannels can be identified from a time-gated reflection matrix constructed by the images of backscattered waves measured at the time-gating window for the trajectory (3). We observe that the enrichment of backscattering intensity at $\tau_0$ leads to the focusing of the snake-like multiple-scattered waves to the target object.

To support our claim of wave focusing inside the scattering medium, we performed numerical simulations using the finite-difference time-domain (FDTD) method. Since the FDTD method can compute electromagnetic wave propagation in arbitrary media (see Supplementary Section II for details), it provides a perfect platform to investigate the effect of wave interference in the middle of the scattering medium. We numerically prepared a scattering medium whose thickness is 30 μm, and scattering and transport mean free paths are $l_s$ = 7.5 μm and $l_t$ = 30 μm, respectively. The scattering medium was located between the two dashed lines in Fig.1c, and a target with width of 10 μm was placed at the depth of $z_s$ = 21 μm from the surface of the scattering medium. The transmittance of the target was about 50 %. An ultra-short pulse was sent to the medium, and its propagation inside the medium was calculated over time. The temporal pulse width of the incident wave was set to 17 fs, which becomes the minimum width of the time-gating window. Light waves were sent through a 40 μm-width window in the middle of the scattering medium and backscattered waves were detected through the same window. When we sent an incident wave of an arbitrary phase pattern, we observed that the waves were diffused laterally and reduced in intensity with increasing depth (Fig. 1c). The backscattered waves from the target were also diffused on their way to the surface (Fig. 1e). We then identified an eigenchannel that maximizes the intensity of backscattered waves at the gating time $\tau_0$ and computed the propagation of the wave when the incident light was coupled to the eigenchannel of the maximum eigenvalue. As shown in Fig. 1d, most of the waves were guided to the target and only few were diffused out. In this case, the light energy delivery has been enhanced by 2.7 times in comparison with random input and the backscattered waves experienced less diffusion than the random input case so that they were better captured at the detection window (Fig. 1f).

## Results

### Sample preparation

For the experimental demonstration of enhancing light energy delivery to the target object embedded within a scattering medium, we prepared a test sample shown in Fig. 2a. We fabricated a target object wherein a silver disk with the diameter of 10 μm was coated on a 1 μm-thick transparent sheet of Poly(methyl methacrylate) (PMMA). Figure 2c shows the reflectance image of this target exposed to free space. The thickness of the silver disk was about 30 nm, which was thin enough that the transmittance measured right behind the disk was about 62.6 % (see Supplementary Section III for the transmittance measurement of the silver disk). With the known transmittance, we could directly measure the light intensity at the target from the transmission measurement. On the top of the target object, we placed a scattering layer consisting of randomly dispersed 1 μm-diameter polystyrene beads in Polydimethylsiloxane (PDMS). The scattering parameters of this layer were $l_s$ = 48.5 μm and $l_t$ = 190 μm at a wavelength of 780 nm. The thickness of the scattering layer, or the depth of the target $z_s$, was varied from 1.9 $l_s$ to 6.8 $l_s$. Another scattering layer with similar scattering parameters was placed at the bottom to embed the silver disk in the middle of the scattering medium. The effect of the thin sheet of PMMA hosting the silver disk was negligible as its index was almost the same as that of the scattering layer and its thickness was much smaller than that set by the width of the time-gating window.

As a representative sample, we considered a silver disk embedded at $z_s$ = 331.7 μm (6.84 $l_s$ and 1.75 $l_t$) from the surface of the scattering medium. Time-gating reflection measurement was performed by using low-coherence interferometry (see Supplementary Section IV for detailed experimental setup). A random phase pattern was written on the wavefront of the output beam from a Ti:Sa laser (center wavelength: 780 nm, pulse width: 52 fs) illuminating the sample, and the total intensity of backscattered waves was measured as a function of flight time (Fig. 2b). A strong peak appeared at the depth corresponding to the surface of the scattering medium due to the index mismatch between immersion water of the objective lens and the medium. We set this reflection as a reference point and assigned its flight time to zero. The width of this peak was measured to be 7.8 µm, which is equal to the time-gating width of our imaging system determined by the pulse width of the light source. A small peak indicated by a black arrow appeared at $\tau_0$ = 3.1 ps, which corresponds to the backscattered waves originating from the silver disk. Note that the peak was broadened in comparison with that from the surface because of multiple light scattering. There were plenty of backscattering signals arriving at the other flight times, and their integral was 86 times larger than the signal at $\tau_0$. For this reason, the target was completely invisible in the steady-state reflectance imaging in which backscattering from all time of flight were added together. The target was invisible even in the angular compounding imaging gated at $\tau_0$ (Fig. 2d) as the backscattered waves got spread on its way back to the detector. Moreover, the target was not resolved even via CASS microscopy[21] (Fig. 2e), which selectively collects single-scattered waves for image reconstruction. This indicates that the target was embedded so deep that the signal strength of single-scattered waves was too small to be resolved.

Next, we measured a time-gated reflection matrix at $\tau_0$. To cover all the input free modes in the system, a transverse wavevector of incidence wave $\vec{k}^i = (k_x^i, k_y^i)$ was scanned by writing linear phase ramps of various orientations and slopes on the spatial light modulator (SLM). The angular scanning range spanned up to the numerical aperture of 0.4, and the number of incidence wavevectors was 1,600 to cover the orthogonal free modes for the view field of $40 \times 40 \ \mu m^2$. Figure 3b shows some of the representative amplitude maps of the backscattered waves, $E_o(\vec{r}_o; \vec{k}^i, \tau_0)$, acquired for each $\vec{k}^i$. Here the objective focus was set to the target depth so that $\vec{r}_o = (x, y)$ corresponds to the spatial coordinates at the target plane where the silver disk was placed. A set of the complex field maps $E_i(\vec{r}_i; \vec{k}^i, \tau_0)$ of the phase ramps written on the SLM was used as an input basis (Fig. 3a). Here, $\vec{r}_i = (x, y)$ is the same as $\vec{r}_o$ because the focus of illumination was matched to that of the collection.

To identify time-gated reflection eigenchannels, we constructed a time-gated reflection matrix $R(\vec{r}_o; \vec{r}_i, \tau_0)$ whose elements consist of the complex amplitude at a detection point $\vec{r}_o$ and a specific flight time $\tau_0$ for the illumination of a unit-amplitude incident wave at a position $\vec{r}_i$. For this purpose, the set of measured images in Fig. 3b were reshaped to form a matrix $M_o(\vec{r}_o; \vec{k}^i, \tau_0)$ by converting individual images into columns of the matrix. Similarly, a matrix $M_i(\vec{r}_i; \vec{k}^i, \tau_0)$ was constructed from the input basis. Next, the time-gated reflection matrix was constructed by the matrix multiplication, $R(\vec{r}_o; \vec{r}_i, \tau_0) = M_o(\vec{r}_o; \vec{k}^i, \tau_0) M_i(\vec{r}_i; \vec{k}^i, \tau_0)^{-1}$. Figure 3c shows the amplitude map of the measured $R(\vec{r}_o; \vec{r}_i, \tau_0)$ for the scattering sample used in Fig. 2a.

To identify the eigenchannels of this time-gated reflection matrix, we performed a singular value decomposition:

$$R = U\Lambda V^\dagger. \qquad (1)$$

Here, V and U are unitary matrices whose columns contain eigenchannels at the input and output planes, respectively. $\Lambda$ is a diagonal matrix whose diagonal elements called singular values are real non-negative. The singular value $\sigma_j$ was sorted in the descending order with respect to the eigenchannel index $j$. Each singular value and its squared value, or eigenvalue, correspond to reflectance of amplitude and intensity, respectively, for the associated time-gated eigenchannel. The squared singular value $\sigma_j^2$ is plotted as blue dots in Fig. 3d. The plot was normalized by the reflectance of a random incident wave.

**Coupling to the time-gated reflection eigenchannels**

To inject light to individual time-gated eigenchannels, we shaped incident waves as those identified from the columns of V. For example, we converted the $j^{th}$ column of V into a two-dimensional complex amplitude image in the $\vec{r}_i$ plane, which is the $j^{th}$ eigenchannel with singular value of $\sigma_j$. Since the SLM used in the experiment can control only the phase of the incident wave, only the phase map of the eigenchannel could be written on the SLM. Still, the accuracy of wavefront shaping measured by the normalized cross correlation between the output eigenchannels identified from the matrix and those experimentally generated was 0.8 or larger. Experimentally measured reflectance of some of the representative eigenchannels are shown as red dots in Fig. 3d. The plot was normalized by the reflectance of the experimentally generated random incident waves. The reflectance of the smallest $j$ was almost four times larger than the random input. We could observe the monotonic decrease of reflectance with the increase of $j$, similar to the prediction from singular values of the measured matrix (blue dots) and their phase-only correction (green dots). Since the coupling of incident light to the time-gated eigenchannels led to the changes in the reflectance at the flight time where the matrix was measured, the temporal response of the reflected wave is expected to be modified. As shown in Fig. 3e (red curve), the intensity of backscattered waves was increased by a factor of 3.8 at the target flight time $\tau_0 = 3.1\ ps$ indicated by a black arrow for eigenchannel index of $j = 1$ in comparison with random input (blue curve). Interestingly, the temporal response for the other flight times remained almost the same. This suggests that eigenchannels for one specific flight time are completely uncorrelated to those obtained at the other flight times.

We also investigated the way the steady-state eigenchannels[9,16] that we previously reported affect to the temporal response of the backscattered waves. We experimentally measured the reflection matrix of the same sample using a continuous wave (see Supplementary Section V for the reflection matrix measurements using continuous waves). This was simply realized by turning off the mode-lock of the Ti:Sa laser. After identifying the eigenchannels of the steady-state reflection matrix, we shaped their wavefront using SLM and measured their temporal response. As shown in Fig. 3e (green curve), the intensity of the backscattered waves was increased at all the flight times for the eigenchannel index of $j = 1$ in comparison with the random input. This is a clear distinction from the time-gated reflection eigenchannels in which the intensity was enhanced only at the target flight time.

Next, we experimentally proved that the control of temporal response of the backscattered waves led to an efficient light energy delivery to the embedded target. We recorded the images of backscattered waves in Fig. 4d-f when incident waves were shaped and coupled to random inputs (Fig. 4a), the steady-state eigenchannels (Fig. 4b), and the time-gated eigenchannels (Fig. 4c), respectively. The images were recorded at the time-gating window centered at $\tau_0$ for the view field of $40\times40\ \mu m^2$. The first 10 highest eigenchannels were averaged for Fig. 4e and 4f, and 50 random phase patterns were used to obtain Fig. 4d. In accordance with the observations in Fig. 3, the reflection intensity of the time-gated eigenchannels is larger than those of the random input and the steady-state eigenchannels. The total intensity in Fig. 4f was 3.33 and 1.77 times larger than those in Fig. 4d. and Fig. 4e, respectively. The enhancement factor is smaller than that shown in Fig. 3, which was 3.8 times, as these images in Figs. 4e and 4f are the average of the first 10 eigenchannels.

For the unambiguous proof that the incident light was focused to the target in the case of time-gated eigenchannel, we measured the transmission images through the scattering medium. For the direct observation of light energy reaching the silver disk, we removed the scattering layer located below the target and measured the transmission images. The transmission images were recorded in the steady-state setting in which signals from all the flight times were recorded together, and the view field of the transmission image was $180\times180\ \mu m^2$. Since the transmittance of the silver disk is 62.6 %, we can directly determine the intensity of light delivered to the target from the measured intensity in the transmission side. In a separate experiment where the bottom scattering layer was in place, we confirmed that the existence of the bottom layer hardly affected the overall behavior, such as the temporal responses of the eigenchannels, their enhancements of reflection intensity and depth-dependent transmission enhancement (see Supplementary Section VI for the effect of the bottom scattering layer). In the case of random input, the silver disk appeared dark (Fig. 4g) in comparison with the surrounding area, which is due to its 62.6 % transmittance. The disk appeared dark in the transmission image of the steady-state eigenchannels as well (Fig. 4h), suggesting that they were not effective either in delivering light energy to the target. On the other hand, the intensity at the disk was higher than the surrounding area in the transmission image of the time-gated eigenchannels (Fig. 4i), which is a direct evidence that more light energy was delivered to the target than when using the random input. Note that the transmitted waves were heavily diffused after experiencing multiple scattering through 1.75 times the $l_t$. The $40 \times 40 \ \mu m^2$ square pattern of illumination was completely smeared out and the width of the overall spatial distribution of light intensity was broadened to 129 μm in terms of the full-width at half-maximum.

To quantify the enhancement of light energy delivery, we obtained the radial intensity profiles of the transmission images with respect to the center of the silver disk (Fig. 5a). To reduce the speckles, the intensity distribution of the transmission images was averaged along the circular equidistant lines from the center of the silver disk. The profile of the random input (blue curve) exhibited a dip at the target position as expected from the dark circular spot in Fig. 4g. A similar dip appeared at the profile of the steady-state eigenchannels. In contrast, the profile of the time-gated eigenchannels has a bump at the center, which is due to the bright circular spot in Fig. 4i.

What we obtained in Fig. 5a are the intensity profiles of the transmitted waves through the silver disk. To obtain the profiles of intensity at the front surface of the target, we need to account for the transmittance of the silver disk. To do this, we recorded the intensity profile of the random input without the target (blue curve in Fig. 5b). Then, the transmittance map of the silver disk was obtained by dividing the profile of the random input with the target by that without the target. By dividing the transmittance map to the line profiles in Fig. 5a, we obtained the intensity profiles at the front surface of the target (Fig. 5b). The intensity profile of the time-gated eigenchannel (red curve in Fig 5b) shows a clear peak at the target region when compared with the profile of the random input. This is the direct proof that the random diffusion was suppressed and the light energy was indeed better focused to the target. The background diffusion outside the target region shows little change because the transmission images were recorded in the steady-state setting in which signals from all the flight times were recorded together. The enhancement factor of light energy delivery to the target is 4.23 times at the center of the disk for the average of the first 10 time-gated eigenchannels. For the eigenchannel index of $j = 1$, the enhancement factor was measured to be 4.96 times in comparison with the random input (see Supplementary Section VII for the image of each eigenchannel). The total intensity curve of the steady-state eigenchannels (green curve in Fig 5b) shows no peak but enhancement of 1.5 times throughout the whole region. This again shows that the steady-state eigenchannels is not good at target specific energy delivery when the target is deeply embedded within scattering media.

We also investigated the dependence of light energy delivery on the depth of the target. In Fig. 5c, the radial intensity profiles of the time-gated eigenchannels are shown for the depths of 1.87 $l_s$, 3.45 $l_s$, 5.21 $l_s$, and 6.84 $l_s$ after accounting for the transmittance of the silver disk. Clear enhancement was observed for all depths. Note that the profiles outside the target area show the degree of wave diffusion. In the case of $z_s = 1.87$ $l_s$, the 40 μm-width of the illumination pattern was relatively well preserved at the baseline due to the weak diffusion. The pattern of the illumination beam was broadened and became invisible with increasing target depth. From the intensity at the target in profiles of Fig. 5c, the enhancement factor of the light energy delivery was obtained (Fig. 5d). First, the enhancement factor increased to 7.35 as the depth increased to a value of 5.21 $l_s$. This is the regime where the turbidity of the scattering medium is not large enough for the phase only eigenchannels to be fully effective. For instance, $z_s$ = 3.57 $l_s$ is about 0.88 times the transport mean free paths, so the direction of propagation of the light waves is not fully randomized. If an SLM capable of controlling both amplitude and phase is used, the enhancement at this regime can be increased. Second, the enhancement factor decreased as the target depth increased beyond $z_s$ = 5.21 $l_s$. In this regime, the action of injecting light to the time-gated eigenchannel becomes less effective as the target depth increases. This is mainly because the fraction of the backscattered waves that have interacted with the target decreases, which means that the enhancement of the intensity at a target flight time is less correlated with the enhancement of light energy delivery to the target. The fraction measured from experiment was 0.75 and 0.41 for the depth of 5.21 $l_s$ and 6.84 $l_s$, respectively. The relative decrease in this value is comparable to the relative decrease in the enhancement from 7.35 to 4.23.

**Light energy delivery through a rat skull**

We demonstrated focusing of light energy to the target located below biological tissues. As shown in Fig. 6a, we placed a target under a skull from a three-day-old Sprague Dawley rat and performed experiments in the epi-detection geometry. The thickness of the skull was approximately 340 μm. Two types of the target were prepared: (1) a 10 μm-diameter silver disk (a target with the same properties of the target in Fig. 2) and (2) ten sparsely spread 3 μm-diameter silver disks all with 30 nm thickness. In Fig. 6b, the temporal response of the backscattered waves is shown for the time-gated eigenchannel of $j$ = 1 (red) and random input (blue) in the case of the target (1). The skull shows more complex temporal response of the backscattered waves than the artificial scattering samples due to its structural inhomogeneity. The intensity of backscattered waves was increased by a factor of 33.6 at the target flight time $\tau_0 = 3.2\ ps$ in comparison with random input. The targets were completely invisible under the skull in both targets (Figs. 6c, d) when imaged by CASS microscopy. We recorded the images of the transmitted waves when incident waves were shaped and coupled to random inputs (Figs. 6e, f) and the time-gated eigenchannels (Figs. 6g, h), respectively, for target (1) and (2). The maximum energy delivery enhancement to the target was 12.4 in comparison with the random input for target (1), which is an experimental record. This result shows the potential of our method to efficiently focus energy through a skull, which is the main barrier for optical imaging and stimulation of nervous systems. Since the proposed method works in the epi-detection geometry, it is non-invasive and therefore readily applicable to the real practices.

**Discussion**

While most previous studies have been concerned with the control of multiple-scattered waves for the targets located outside of a scattering medium, we presented a method to control them inside the medium. This is particularly important for *in vivo* and *in situ* applications because the target objects that we would like to visualize or control are mostly embedded within disordered environments. Furthermore, the proposed method works non-invasively since it operates in the epi-detection geometry. In our method, we performed time-gated imaging to gain information about the trajectories of the light waves inside the scattering medium. A specific flight time was chosen at which a relatively large fraction of multiple-scattered waves have interacted with the target object, and a reflection matrix was recorded at this selected flight time to identify its reflection eigenchannels. By injecting light to the eigenchannels with large reflectance, we could alter the diffusive behavior of multiple-scattered waves and focus them to the targeted object. In doing so, we could enhance the light energy that reaches the target by a factor of more than 10 in comparison with the random input. The proposed method is so general that it can potentially be extended to non-destructive applications dealing with wave propagation in complex media, such as structural engineering, petroleum engineering, and forensic science as well as biomedicine.

Our method can particularly be useful for the scattering media such as biological tissues whose spatial refractive index variation is relatively small. In general, the linear relationship between the flight time and flight depth of the backscattered waves is not valid when strong multiple scattering occurs. However, this relationship can be preserved to a certain extent in these scattering media where forward scattering is dominant. For the choice of a flight time set by the target depth, the multiple-scattered waves whose trajectories resemble a 'snake' are detected at the finite width of the time-gating window in addition to the single-scattered waves that travel straight to the target. In our study, we controlled these snake-like multiple-scattered waves to focus them to the target object. As a working condition for the proposed method, the target object needs to have a higher reflectance than the surrounding scattering media. In fact, this is the case for any optical methods exploiting reflectance as an intrinsic source of contrast. For biological specimens, structures like myelin can be a good target since lipid in the myelin has a larger refractive index than the surrounding tissue. It will be an interesting future work to efficiently deliver light energy to these biological targets and stimulate their function. Other potential target objects may be optically chargeable implantable medical devices[22] and tumors treated with gold particles for photothermal therapy.

The proposed method worked for the targets that were embedded too deep to be resolved by optical imaging. This was made possible because we used snake-like but multiple-scattered waves whose intensity is much stronger than the single-scattered waves. Note that the imaging depth of the deep-tissue optical imaging modalities such as optical coherence microscopy and CASS microscopy are shallow because they rely on single-scattered waves for the image reconstruction. However, the proposed method certainly has limitations in its working depth. The fraction of the backscattered waves that have interacted with the target decreases with increasing depth, and the increase in the intensity of the selected flight time does not necessarily enhance light energy delivery to the target. The finite detector dynamic range can pose technical limits because the signals from other flight times can waste most of the detector dynamic range. For a better translation of the proposed concept to real applications, the speed of the operation needs to be improved. A liquid crystal based spatial light modulator was used in the present study, but the use of high-speed beam shaping devices such as digital micromirror devices can be used for operation times well below one second. The real-time feedback control method that we proposed in the steady-state experiments[23,24] can be implemented in the time-gated experiments, which can potentially speed up the operation and keep pace with fluctuations of the samples. With these improvements, the proposed method can lead to further improvement of the working depth of *in vivo* optical imaging, sensing, and stimulation.

**Acknowledgements**

This research was supported by IBS-R023-D1, and the Global Frontier Program (2014M3A6B3063710) through the National Research Foundation of Korea (NRF) funded by the Ministry of Science, ICT & Future Planning. It was also supported by the Korea Health Technology R&D Project (HI14C0748) funded by the Ministry of Health & Welfare, Republic of Korea.

**Author contributions**

W.C., S.J., S.K. and Y.-R.L. conceived the experiment. S.J. and Y.-R.L. carried out the measurements and analyzed the data with W.C.. W.C.* and Y.-R.L. performed the theoretical study and supported interpretation of the data. Y.-S.L. assisted in the design of the optical set-up. J.H.H. prepared biological tissues. J.P. and H.P. provided silver disks. S.J., Y.-R.L. and W.C. prepared the manuscript. All authors contributed to finalizing the manuscript. W.C. and W.C.* refer to Wonshik Choi and Wonjun Choi, respectively.

**Materials & Correspondence**

To whom correspondence and requests for materials should be addressed: Wonshik Choi, wonshik@korea.ac.kr.

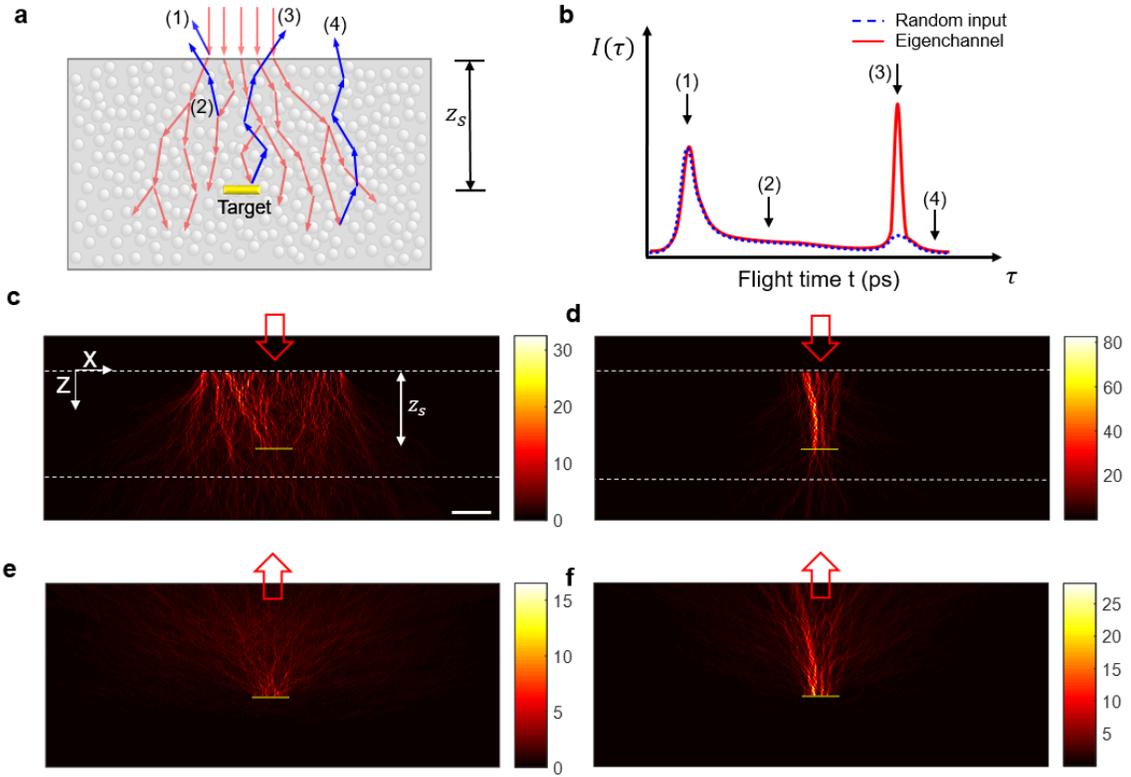

**Figure 1. Wave propagation within scattering media depending on the time of flight. a,** Schematic trajectories of wave propagation for a scattering medium with an embedded target. (1) back-reflection at the surface. (2), (4) multiple-scattered waves that have not interacted with the target object, and (3) waves that have interacted with the target object. **b,** An exemplary intensity profile of backscattered waves as a function of flight time. Flight times indicated by (1-4) correspond to those trajectories shown in **a**. By coupling light to the time-gated reflection eigenchannels at the flight time (3), the intensity of the backscattered waves at the selected time-gating window can be increased (red curve). The blue dotted curve shows the intensity profile for the random input. **c-f,** Wave propagation calculated from FDTD simulation. The scattering medium is located between the dashed lines, and a target object is located at the depth of $z_s$ = 21 μm. For the random input, the intensity maps of the forward (**c**) and backward (**e**) propagating waves were accumulated during the round trip. **d** and **f**, The same as **c** and **e**, respectively, but for the case of a time-gated reflection eigenchannel with the largest singular value. Scale bar, 10 μm. Color bars, intensity in arbitrary units, but all at the same scale.

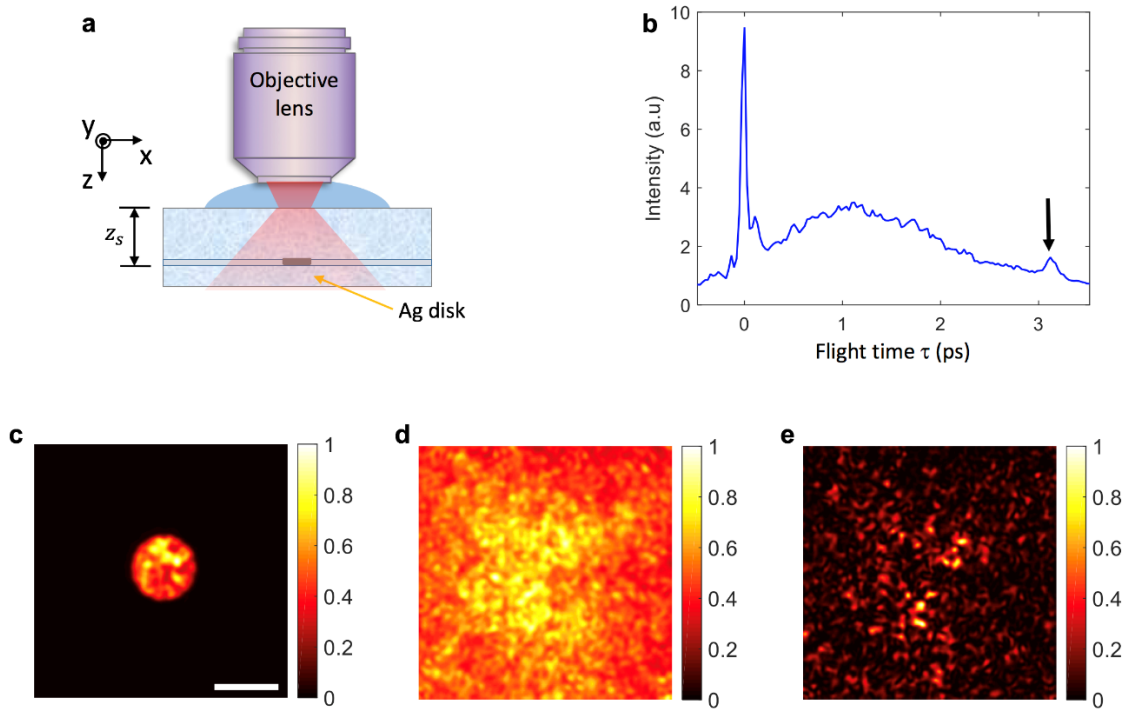

**Figure 2. Preparation of a target embedded in a scattering medium. a,** A schematic diagram of the sample. A 10 μm-diameter and 30 nm-thick silver disk coated on a 1 μm-thick transparent PMMA was used as a target object. The target layer was embedded between scattering layers made of PDMS with randomly dispersed 1 μm-diameter polystyrene beads. $z_s$: the thickness of the upper scattering layer, and therefore the depth of the target. **b,** Intensity of the backscattered waves as a function of the flight time for the illumination of random pattern. The peaks at $\tau = 0$ ps and $\tau_0 = 3.1$ ps correspond to the back-reflections at the surface and from the target located at $z_s = 331.7$ μm respectively. **c,** A reflectance image of the target object with no scattering layer on the top, i.e. $z_s = 0$. **d,** Angular compounding image of the sample after the time-gated detection. **e,** Reconstructed image of the sample using CASS microscopy. Scale bar, 10 μm. Color bars, normalized intensity.

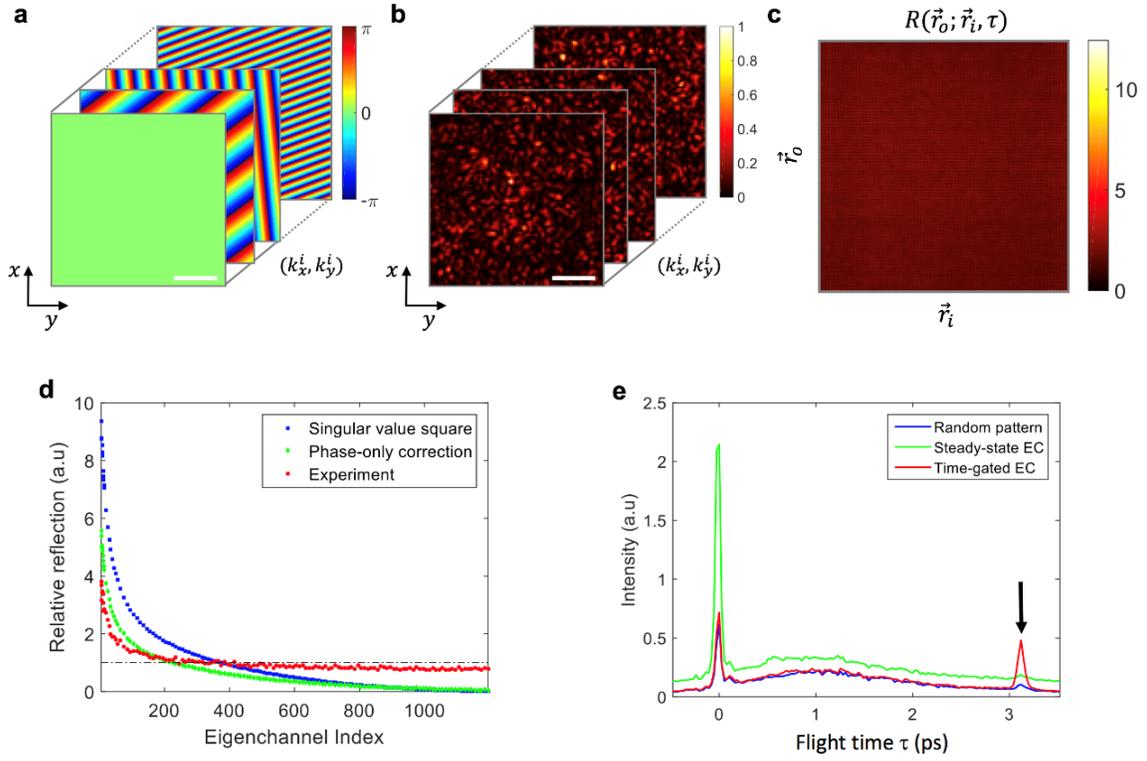

**Figure 3. Experimental measurement of a time-gated reflection matrix and coupling light to its eigenchannels. a,** Linear phase ramps of various orientations and slopes written on the spatial light modulator. Scale bar, 10 μm at the sample plane. Color bar, phase in radians. **b,** Time-gated amplitude maps of the backscattered waves for various incident wavevectors taken at $\tau_0$ = 3.1 ps. Phase maps were acquired simultaneously, but are not shown here. Color bar, normalized amplitude. **c,** A time-gated reflection matrix $R(\vec{r}_o; \vec{r}_i, \tau_0)$ constructed from 1,600 complex field maps shown in **a** and **b**. Column and row indices are $\vec{r}_i$ and $\vec{r}_o$, respectively. Color bar, amplitude in arbitrary units. **d,** The square of singular values calculated from the singular value decomposition of the time-gated reflection matrix (blue dots). Green dots: the square of singular values after accounting for the phase-only control of the eigenchannels. Red dots: the total intensity of the backscattered waves at the target flight time for the case of experimentally coupling light to individual time-gated reflection eigenchannels. Plots were normalized with respect to random input. **e,** The temporal response of the backscattered waves for the time-gated eigenchannel of $j$ = 1 (red), random input (blue), and steady-state eigenchannel of $j$ = 1 (green).

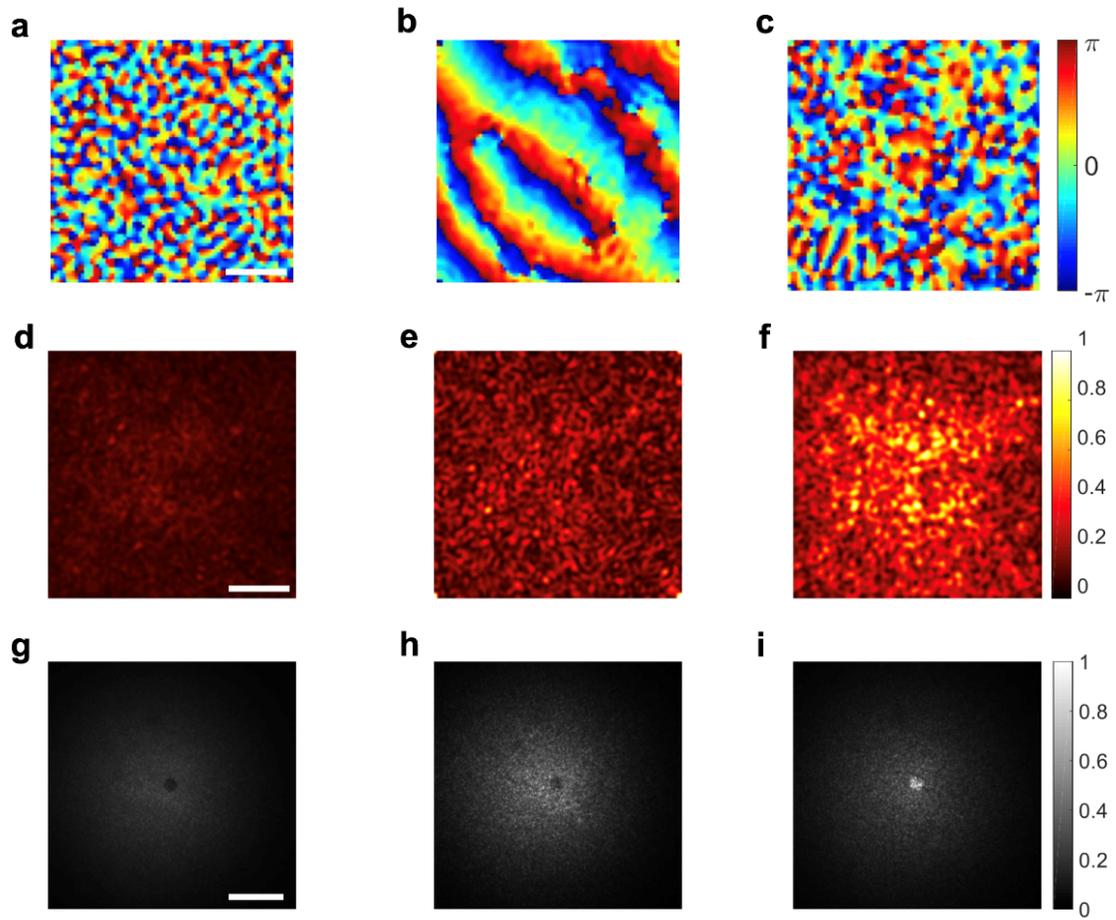

**Figure 4. Reflection and transmission images of the time-gated reflection eigenchannels.** The top row: the phase maps of illumination for the random input (**a**), the steady-state eigenchannel of $j = 1$ (**b**), and the time-gated eigenchannel of $j = 1$ (**c**). Scale bar, 10 μm. Color bar, phase in radians. **d-f**, the intensity maps of the backscattered waves at $\tau_0 = 3.1$ ps for the illuminations of **a-c**, respectively. Scale bar, 10 μm. Color bar, intensity normalized by the maximum intensity at **f**. **g-i**, Intensity maps of transmitted waves for the illuminations of **a-c**, respectively. Scale bar, 40 μm. Color bar, intensity normalized by the maximum intensity at **i**. For the case of reflection and transmission images of eigenchannels, eigenchannels with $j = 1$ to $j = 10$ were averaged.

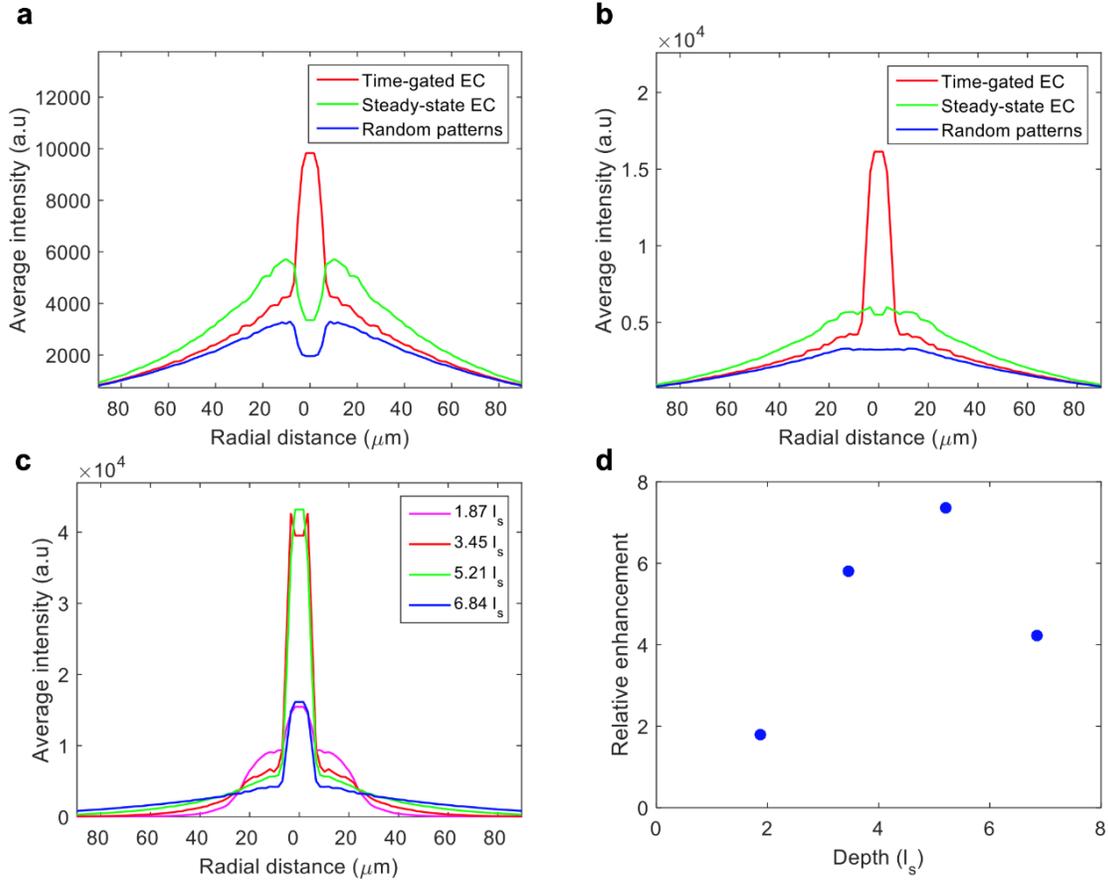

**Figure 5. The enhancement factor of light energy delivery for the time-gated reflection eigenchannels. a,** Radial intensity profiles of the transmission images. Red, green, and blue curves correspond to the time-gated eigenchannels, steady-state eigenchannels, and random inputs, respectively. **b,** Radial intensity profile at the front surface of the silver disk after accounting for the transmittance of the silver disk. **c,** Radial intensity profiles of the time-gated eigenchannels at the front surface of the silver disk depending on the depth of the target. Magenta, red, green and blue curves are from $z_s$ = 1.87 $l_s$, 3.45 $l_s$, 5.21 $l_s$, and 6.84 $l_s$, respectively. **d**, Enhancement factor of light energy delivery for the time-gated eigenchannels depending on the depth of the target.

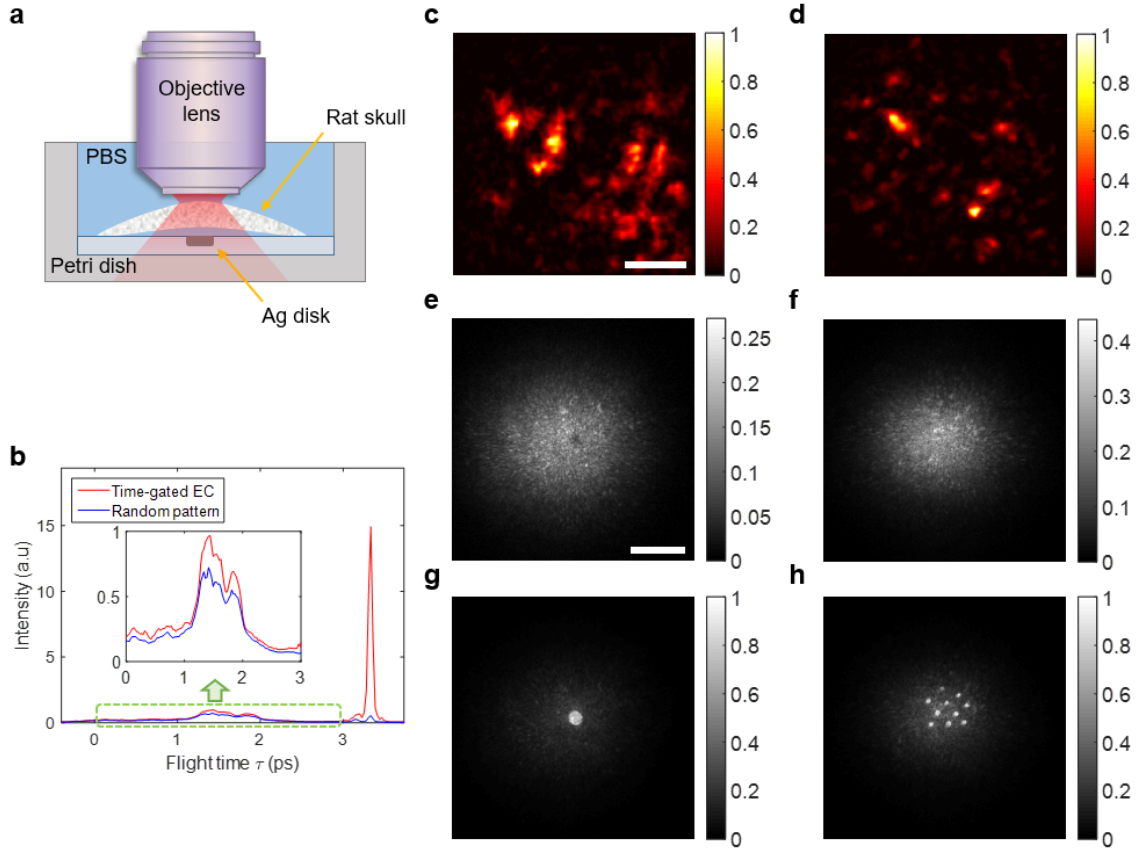

**Figure 6. Demonstration of enhanced light energy delivery through a rat skull. a,** Sample geometry. Targets were covered by approximately 340 μm-thick rat skull extracted from a three-days-old Sprague Dawley rat. Two different types of targets were prepared: (1) a 10 μm-diameter silver disk and (2) ten sparsely spread 3 μm-diameter silver disks all with 30 nm thickness. **b,** The temporal response of the backscattered waves for the time-gated eigenchannel of $j = 1$ (red) and random input (blue) in the case of the target (1). Inset: zoom-in plot for the range marked by the green dashed box. **c,** and **d,** The CASS microscopy images of the sample with target (1) and (2), respectively. Scale bar, 10 μm. Color bar, normalized intensity. **e,** and **f,** Intensity maps of transmitted waves for the cases of random inputs in the order of target (1) and (2). Scale bar, 40 μm. **g,** and **h,** Intensity maps of transmitted waves averaged for the time-gated eigenchannels of $j = 1$ to $j = 50$ in the order of target (1) and (2). Color bars, intensity normalized by the maximum intensity in **g** and **h** for each target.